\documentclass[pre,nofootinbib,superscriptaddress,onecolumn,preprintnumbers]{revtex4}
\usepackage{graphicx}
\usepackage{amsmath,amssymb}
\usepackage{color}

\renewcommand{\eqref}[1]{Eq.(\ref{#1})}

\newcommand{\hz}{\mathcal{H}_0}

\begin{document}
\title{Running Hubble constant: evolutionary Dark Energy}

\author{G. Montani}
\email{giovanni.montani@enea.it}
\affiliation{ENEA, Nuclear Department, C. R. Frascati, Via E. Fermi 45,  Frascati, 00044 Roma, Italy}
\affiliation{Physics Department, ``Sapienza'' University of Rome,  P.le Aldo Moro 5, 00185 Roma, Italy}

\author{N. Carlevaro}
\affiliation{ENEA, Nuclear Department, C. R. Frascati, Via E. Fermi 45,  Frascati, 00044 Roma, Italy}

\author{M.G. Dainotti}
\affiliation{Division of Science, National Astronomical Observatory of Japan, 2-21-1 Osawa, Mitaka, Tokyo 181-8588, Japan}
\affiliation{The Graduate University for Advanced Studies (SOKENDAI), Shonankokusaimura, Hayama, Miura District, Kanagawa 240-0115}
\affiliation{Space Science Institute, Boulder, CO, USA}

\begin{abstract}
We discuss an evolutionary dark energy model, based on the presence of non-equilibrium effects on the dark energy constituents, which are described via a bulk viscosity contribution. We implement the proposed dynamics by the analysis of the 40-bins Type Ia Supernovae (SNe) Pantheon sample data, in order to outline the existence of a running Hubble constant with the redshift. Via a fitting procedure, we determine the value of the additional parameter that our model possesses with respect a standard $\Lambda$ Cold Dark Matter ($\Lambda$CDM) scenario. As important result, the evolutionary dark energy proposal seems more appropriate to describe the binned SN analysis with respect to the $\Lambda$CDM Hubble parameter, i.e. a non running value for the Hubble constant over the bins.
\end{abstract}

\maketitle

In recent years, the so-called \emph{Hubble tension} \cite{DiValentino:2021izs,Dainotti2021apj-powerlaw, Dainottigalaxies10010024,desimone2024doubletcosmologicalmodelschallenge} has attracted increasing interest, as its resolution could shed light on some of the most puzzling aspects of the late Universe, such as the true nature of dark matter or dark energy. The crux of the tension lies in the nearly $5\sigma$ discrepancy between two primary methods of determining the Hubble constant, $H_0$: the SH0ES Collaboration measurements \cite{riess2022apjl,Brout:2022vxf}, which rely on relatively low-redshift sources (mainly Type Ia Supernovae (SNe)), and the Planck Satellite Collaboration \cite{Planck:2018vyg}, which uses Cosmic Microwave Background (CMB) photons.

Many models have been proposed to explain the Hubble tension, most of which suggest late-Universe modifications to the standard $\Lambda$ Cold Dark Matter ($\Lambda$CDM) model, see innovative propositions \citep{Dainotti2021apj-powerlaw, Dainottigalaxies10010024,desimone2024doubletcosmologicalmodelschallenge} (see also the recent analysis in \cite{2025arXiv250111772D}), and the reviews \cite{DiValentino:2021izs,hu-wang03,2018PhRvD..97d3513D,2018PhRvD..97d3528D,schiavone_mnras,2024PDU....4401486M,deangelis-fr-mnras,2024PDU....4601652E,2024arXiv240415977M}. In the context of late-Universe dynamics, the tension becomes evident in the divergence between data derived from Cepheid calibrated SNe and Baryon Acoustic Oscillations (BAOs). In fact, SNe yield a higher $H_0$ value than that derived from BAO measurements. BAOs, in contrast, tend to produce a $H_0$ value consistent with that measured by Planck, although there are indications that this analysis may not be entirely model-independent \cite{DESI:2024mwx}. This discrepancy between SN and BAO data yielded proposals of early-Universe modifications to the physics underlying cosmic expansion, such as the \emph{early dark energy} hypothesis \cite{Kamionkowski:2022pkx} (see also \cite{2021PDU....3200837N}). Some proposals have even suggested that both early- and late-time modifications to the standard cosmological model may be necessary to fully account for the observed differences in $H_0$ measurements \cite{2023Univ....9..393V,2020PhRvD.102b3518V,hu-wang02}. Moreover,  variation of the Hubble constant from the central value obtained by the SN are visible using Gamma Ray Bursts \cite{Dainotti2023ApJ...951...63D,Dainotti2023mnras,Dainotti2023ApJ...950...45D,Dainotti2024Galax..12....4D,2024arXiv241108097B} and Quasars \cite{Bargiacchi2023MNRAS.521.3909B,Lenart} obtained implementing new the statistical assumptions discussed in \cite{Dainotti:2023ebr}. There is actually also a small variation of the values of dark matter if alternative theories, such as varying dark matter, are considered \citep{Dainotti2024PDU....4401428D,naidoo2024PhRvD,Doroshkevich-1}.

While it might seem that our understanding of the Universe is incomplete and requires a set of \emph{ad hoc} conjectures for theoretical consistency, two key observational results stand out, forming the basis for the motivation of this Letter:
\begin{itemize}
    \item[-] A trend appears wherein the value of $H_0$ decreases as the redshift of the sources used for measurement increases. This is observed by comparing SN, BAO, red giant measurements \cite{Anand:2021sum} and CMB measurements (see also \cite{2022PASJ...74.1095D} for gamma-ray bursts studies). This general trend led to the analysis in \cite{Dainotti2021apj-powerlaw} (see also \cite{Dainottigalaxies10010024,hu-wang02,kazantzidis,Krishnan:2020vaf,2024PDU....4401486M}), which introduced the idea of an effective Hubble constant that varies with redshift, using equipopulated redshift bins for SN data. The study outlined a decreasing power-law behavior for $H_0$ as redshift increases. Moreover, the decreasing trend of $H_0$ with redshift was first reported by H0LiCOW \cite{2020MNRAS.498.1420W}. However, subsequent works, including TDCOSMO IV \cite{2020A&A...643A.165B}, have highlighted that the original H0LiCOW results may have underestimated the uncertainties. This raises a key question about whether the observed decreasing behavior of $H_0$ is intrinsic or due to some hidden biases. Indeed, this trend has been the focus of discussion in several studies, such as \cite{Krishnan:2020obg} and \cite{dainotti2023shedding}. However, the nature of this trend is still unclear and its interpretation varies. For instance, in \cite{Krishnan:2020obg} it is attributes the decreasing trend to including Cosmic Chronometers in bins at larger redshifts, which results in a lower $H_0$. Similarly,  \cite{dainotti2023shedding} showed a decreasing trend of $H_0$ consistent within 2 $\sigma$. This raises the possibility that large error bars may prevent us from seeing the presence of an intrinsic trend. A theoretical explanation for this trend may lie in the scalar-tensor (non-minimal) coupling characteristic of $f(R)$ gravity theories in the Jordan frame \cite{Sotiriou-Faraoni:2010,NOJIRI201159,schiavone_mnras,2024arXiv240801410S}. 
    \item[-] Recently, the DESI Collaboration \cite{DESI:2024mwx} (see also, e.g. \cite{PhysRevD.104.023510,dainotti-desi}) has shown that dark energy's contribution evolves in the late Universe. Their data strongly support that the $w_0w_a$CDM model \cite{Chevallier2001,Linder2003,lemos2019} offers a better fit than the standard $\Lambda$CDM paradigm. This observation provides new insights into the notion of vacuum energy in the current Universe, suggesting that it has a dynamic component \cite{Sola2018,Mukhopadhyay--1}. Since the $w_0w_a$CDM model is more of a phenomenological \emph{ansatz} than a precise physical statement, it becomes crucial to explore the possible physical mechanisms that could drive this dynamical behavior.
\end{itemize}

Here, in this spirit, we consider a modified $\Lambda$CDM model that includes an evolving dark energy contribution. We then use the corresponding dynamics to investigate the behavior of the effective (redshift-dependent) Hubble constant, based on the binned SN Pantheon sample analysis performed in \cite{Dainotti2021apj-powerlaw,Dainottigalaxies10010024}. In particular, the evolution of the dark energy density is associated with a non-equilibrium effect, specifically a bulk viscosity contribution in the fluid representation, which affects the fluid constituents during the late Universe expansion.

We begin by considering a flat, isotropic Universe, as suggested by the analysis in \cite{2020MNRAS.496L..91E} (for alternative estimates of the Universe's spatial curvature, see \cite{DiValentino2019-curvature,2021PhRvD.103d1301H}), where the line element takes the form:
\begin{equation}
ds^2 = c^2 dt^2 - a^2(t)\left( dx^2 + dy^2 + dz^2\right)\;, \label{ede1}
\end{equation}
where $t$ denotes the synchronous time and the variables $(x,\, y,\, z)$ represent Cartesian coordinates. The cosmic scale factor $a(t)$ regulates the Universe’s expansion, and by convention, its present-day value $a_0$ is taken to be unity. The evolution of the scale factor is governed by the well-known Friedmann equation which reads \cite{bib:montani-primordialcosmology,KolbTurner1990,Weinberg2008}
\begin{equation}
H^2 \equiv (\dot{a}/a)^2 = (\rho_m+\rho_{de})\chi c^2/3\;,\label{ede2}
\end{equation}
where $\chi$ denotes the Einstein constant, $\rho_m$ is the standard matter (baryonic and dark) energy density contribution, and $\rho_{de}$ is the evolving dark energy density (here the dot indicates differentiation with respect to $t$). Let us now move from the synchronous time to the redshift variable $z \equiv a_0/a - 1=1/a-1$ and in the following we denote by a prime the differentiation with respect to $z$. The matter contribution $\rho_m$ follows the standard behavior $\rho_m' = 3\rho_m/(1+z) $, thus getting
\begin{equation}
\rho_m = \rho_{0m}(1+z)^3\;,\label{ede3}
\end{equation}
where we denote by $\rho_{0m}$ the present-day ($z=0$) value of the matter energy density.

Bulk viscosity is a macroscopic effect that attributes a small dissipative nature to an otherwise perfect fluid, accounting for the difficulty of the microscopic constituents in maintaining thermodynamic equilibrium when the Universe's volume changes \cite{LandauLifshitz1987-fluid,1976AnPhy.100..310I,2015PhRvD..91d3532D,Belinskii1975,Belinskii1977,Belinskii1979,carlevaro-bulk-2005,padmana87,2006GReGr..38..495F,2009CoTPh..52..377M,2011JCAP...09..026G,2013PhRvD..88l3504V,2017EPJC...77..660W}. The net effect of bulk viscosity on the macroscopic dynamics of a fluid introduces a negative pressure term, which alters the standard thermodynamic pressure. Taking into account the dark energy component as a bulk viscous cosmological fluid with an equation of state $p_{de} = w_{de}\rho_{de}$ (where $p_{de}$ is the thermal pressure, $\rho_{de}$ is the energy density and $w_{de}$ is a constant), the continuity equation takes the simple form \cite{bib:montani-primordialcosmology}
\begin{equation}
\rho_{de}' = 3((1+w_{de})\rho_{de} - 3H\xi)/(1+z)\;,\label{ede4}
\end{equation}
where $\xi = \xi (\rho_{de})$ is the bulk (or second) viscosity coefficient assumed, in what follows, as a constant contribution, i.e. $\xi =\bar{\xi} = const.$ The assumption of a bulk viscosity effect, influencing the dark energy component of the Universe is justified by the capability of this energy density to change the cosmological dynamics from a decelerating to an accelerating regime. This transition is reliably associated with a smooth transition phase of the Universe, and the consequent weak deviation from thermal equilibrium can be described via a bulk viscosity coefficient $\xi(\rho_{de})$. The choice to take this coefficient as a constant term is motivated by the limited redshift interval that we are analyzing for our data analysis. In such a range from $z=0$ to $z=2.5$ our choice appears to be a very good approximation, capable of capturing the main features of the proposed phenomenology. A more realistic choice, when the whole Universe dynamics is considered, can be made by requiring that the bulk viscosity coefficient be a power-law in the energy density \cite{Weinberg2008,belinskii1977viscosity}. However, it must be observed that, in different regimes (e.g., close or far from the Big Bang), the associated exponent should have different values.

Using the Friedmann equation to express $H$, we thus rewrite Eq.(\ref{ede4}) as follows:
\begin{equation}
\rho_{de}'= -\frac{3\sqrt{3\chi/c^2}\bar{\xi}}{1+z} \sqrt{\rho_{0m}(1+z)^3 + \rho_{de}}+\frac{3(1+w_{de})}{1+z}\rho_{de}\;. \label{ede5} \end{equation}
Next, recalling that $H_0$ denotes the present day value of the Hubble parameter, we introduce the following natural dimensionless critical parameters:
\begin{equation} 
\Omega_{m0}\equiv \frac{\chi c^2 \rho_{0m}}{3H_0^2}\;,\qquad
\Omega_{de}(z)\equiv \frac{\chi c^2 \rho_{de}(z)}{3H_0^2}\;,\label{ede6}
\end{equation}
and the two basic equations of our modified $\Lambda$CDM model finally take following:
\begin{align}
&H(z)=H_0\sqrt{\Omega_{m0}(1+z)^3 + \Omega_{de}(z)}\;,\label{ede7}\\
&\Omega_{de}'(z) = -\frac{\Omega_*}{1+z} \sqrt{\Omega_{m0}(1+z)^3 + \Omega_{de}(z)}+\frac{3(1+w_{de})}{1+z}\Omega_{de}\;,\label{ede8}
\end{align}
where we have defined $\Omega_*\equiv 3\chi c^2 \bar{\xi}/H_0$, and we clearly have $\Omega_{de}(z=0) = 1 - \Omega_{m0}$. 

Since Eq.(\ref{ede8}) can predict a decreasing behavior for the dark energy parameter as redshift increases, this dynamical proposal is well-suited to address the Hubble tension. The primary aim of this Letter is to compare the effective Hubble parameter, which is denoted henceforth as $\mathcal{H}_0(z)$, against the binned SN Pantheon sample data, as in \cite{Dainotti2021apj-powerlaw,Dainottigalaxies10010024}. The effective Hubble parameter can be defined, according to the analysis in \cite{Krishnan:2020vaf,2022arXiv220113384K,2024arXiv240801410S}, by the following relation which accounts for deviations from a base $\Lambda$CDM model
\begin{align}\label{hzforms}
H(z)= \mathcal{H}_0(z) \sqrt{\Omega_{m0}(1+z)^3 + 1-\Omega_{m0}}\;
\end{align}
We thus obtain, by a direct comparison with Eq.(\ref{ede7}):
\begin{equation}
\mathcal{H}_0(z) \equiv H_0\sqrt{\frac{\Omega_{m0}(1+z)^3 + \Omega_{de}(z)}{\Omega_{m0}(1+z)^3 + 1 - \Omega_{m0}}}\;,\label{ede11}
\end{equation}
which must be coupled with Eq.(\ref{ede8}) for the evolution of the dark energy term.

\begin{figure}[ht!]
\includegraphics[width=0.55\textwidth]{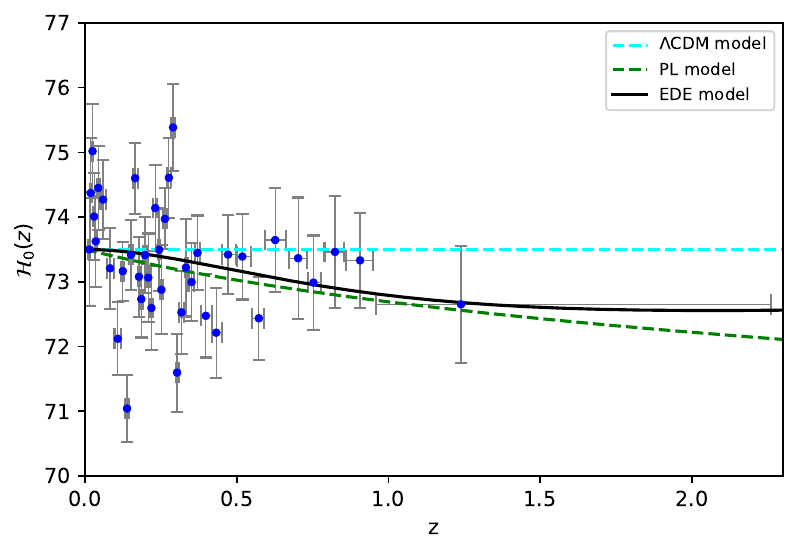}
\caption{Plot of $\mathcal{H}_0(z)$ from Eq.(\ref{ede11}) (black) for the best fit $\Omega_*$ in Eq.(\ref{bestfit}) and the fiducial values $H_0=73.5$ and $\Omega_{m0}=0.298$. The green dashed line represents the profile $H_0(1+z)^{-0.016}$. Blue bullets are $H_0$ data from \cite{Dainotti2021apj-powerlaw} with the corresponding error bars in 1 $\sigma$  and bin widths in the x-axis. We also depict the constant line $\mathcal{H}_0(z)=73.5$ for the base $\Lambda$CDM model (cyan dashed line).}
\label{fig-H0}
\end{figure}

In what follows, we briefly review the data analysis methodology outlined in \cite{Dainotti2021apj-powerlaw,Dainottigalaxies10010024} for constructing the binned analysis of the SN Pantheon sample. For clarity, note that throughout this analysis, the Hubble constant $H_0$ is expressed in km s$^{-1}$ Mpc$^{-1}$. The dataset used for our study is the Pantheon sample \cite{2018ApJ...859..101S}, which compiles 1048 type Ia SNe from various surveys. We divide this dataset into 40 redshift bins, shown in the x-axis of Fig.\ref{fig-H0}, ensuring that each bin contains an equal number of SNe. We have averaged the redshift in each bin: the central point of the bin is already an average of the redshifts of the SNe Ia in that bin. Since we have less SNe Ia at high-z, we have the central value of the redshift shifted at lower redshift. 
The observed distance modulus, $\mu_{\text{obs}} = m_B - M$, is taken from the Pantheon sample, where $m_B$ is the apparent magnitude in the B band (corrected for systematic and statistical errors) and $M$ is the absolute magnitude of a fiducial SN, including corrections for color and stretch. As in previous studies, we average the distance moduli using the models of \cite{2010A&A...523A...7G} (G2010) and \cite{2011A&A...529L...4C} (C2011). The theoretical distance modulus is given by:
$ \mu_{\text{th}} = 5 \log_{10} d_L(z, H_0, ...) + 25$ where the luminosity distance $d_L$ is calculated in megaparsecs (Mpc) under a base $\Lambda$CDM model. We also account for corrections due to the peculiar velocities of host galaxies containing SNe. To quantify the model’s fit, we define the chi-squared statistic for the supernova data as: $\chi^2_{\text{SN}} = \Delta\mu^{T} \cdot \mathcal{C}^{-1} \cdot \Delta\mu$ where $\Delta\mu = \mu_{\text{obs}} - \mu_{\text{th}}$ and $\mathcal{C}$ is the $1048 \times 1048$ covariance matrix, as provided in \cite{2018ApJ...859..101S}. For each redshift bin, we obtain values of $H_0$ by varying $H_0$ while keeping $\Omega_{m0}$ fixed. We specify that we performed MCMC to obtain the best fit for the 40 $H_0$ values within each bin. Regarding the fixed value of $\Omega_{m0}$, we note its variation within the bins will not exceed 2 $\sigma$ of the prior values \cite{Dainotti2021apj-powerlaw}.

The choice of 40 bins strikes a balance between having sufficient data points for accurate fitting and avoiding excessive binning, which could lead to larger uncertainties in $H_0$. Each bin contains around 26 SNe, providing enough data for fitting. Notably, the trend derived for $H_0$ remains unaffected by the initial choice of $H_0$, since the distance moduli in the Pantheon sample are presented with a fixed absolute magnitude $M$. For instance, we use $M = -19.245$, corresponding to $H_0 = 73.5$, as discussed in \cite{Dainotti2021apj-powerlaw}. We fix a given $H_0$ value in the first bin, derive $M$ as the sole free parameter, and fix this value of $M$ for all the other bins, then this procedure does not impact the decreasing trend of $H_0(z)$. 
We remark that while the SH0ES collaboration provides a single $H_0$ value from SN data, in \cite{Dainotti2021apj-powerlaw,Dainottigalaxies10010024} it was argued that subdividing the Pantheon sample into redshift bins reveals a slow decline in $H_0$ with increasing redshift. If the Hubble tension is due to new physics, as proposed here, this effect should influence observations continuously, making an evolving $\hz$ a natural expectation across different redshift domains.

We stress that $\mathcal{H}_0(z)$, as defined in Eq.(\ref{ede11}), is a tool
to investigate if a data set is properly fitted by the $\Lambda$CDM
Hubble parameter or if a deviation, as described in given framework
(here the evolutionary dark energy (EDE)), has to be considered.
In this respect, it is, in our proposal, a function of four free
parameters $H_0$, $\Omega_{m0}$, $\Omega_*$ and $w_{de}$ and all of them are, in principle, available for the fitting procedure. However, in view of the binned scheme, in the following we assign to $H_0$ and $\Omega_{m0}$ the fiducial (in the sense outlined above) values $H_0=73.5$ and $\Omega_{m0}=0.298$. Moreover, we impose the condition $\Omega_{de}'(z=0)=0$ from Eq.(\ref{ede8}). This choice is motivated by the request to remain as much closer as possible to the $\Lambda$CDM model (for which this condition holds always) at sufficiently low $z$-values. In fact, using the standard $q_0$ (deceleration parameter) $j_0$ (jerk parameter) expansion, we get $q_0=-1+3\Omega_{m0}/2+\Omega_{de}'(z=0)/2$ and the modification against a $\Lambda$CDM picture is indeed in the additional derivative of the evolutive dark energy component. This condition allows to safely use the $H_0$ prior as adopted below. We thus obtain the relation
\begin{align}
w_{de}= \frac{\Omega_*}{3(1-\Omega_{m0})}-1\;.
\label{wdeset}
\end{align}
At the same time, we can also estimate the discrepancy from $j_0=1$, proper of the standard $\Lambda$CDM scenario. For our model, implementing the relation above, we get $j_0=1-3\Omega_*\Omega_{m0}/2$. Summarizing, in the present case, the deviation is estimated with respect to a specific reference $\Lambda$CDM model and only the parameter $\Omega_*$ is free for fitting, accounting for the presence of an evolution of the vacuum energy density.

\begin{figure}[ht!]
\includegraphics[width=0.55\textwidth]{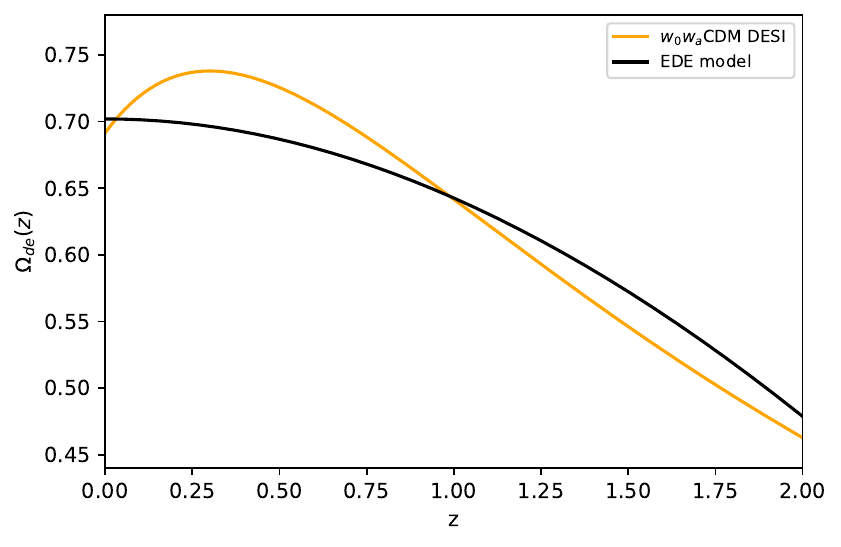}
\caption{Profile of the evolute $\Omega_{de}(z)$ obtained from Eqs.(\ref{ede8}) and (\ref{wdeset}), with the best fit value of $\Omega_*$ in Eq.(\ref{bestfit}) (black line). Orange line represents the evolutive dark energy constrained by DESI.}
\label{fig-Omega-de}
\end{figure}

\begin{figure}[ht!]
\includegraphics[width=0.55\textwidth]{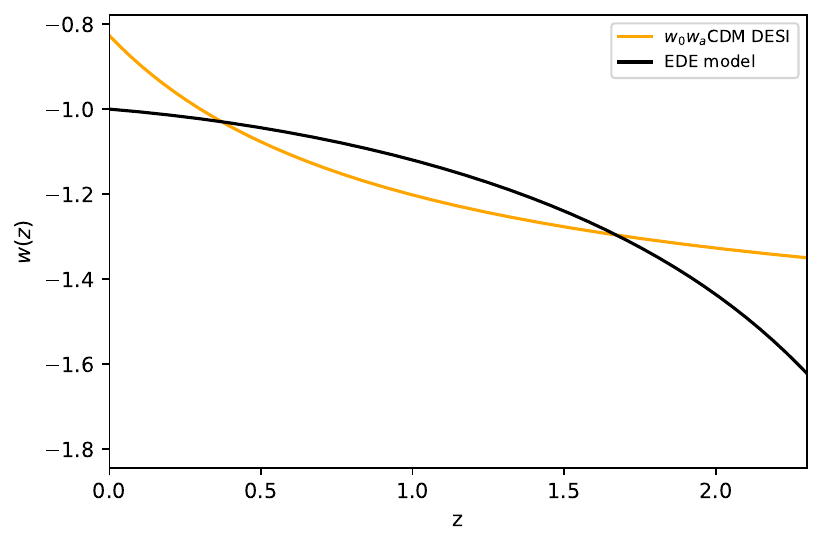}
\caption{Plot of $w_{e\!f\!f}(z)$ from Eqs.(\ref{weff}) (black) and $w(z)$ for the standard $w_0w_a$ parametrization constrained by DESI (orange).}
\label{fig-wz}
\end{figure}

Using the 40 bin distribution of the $H_0$ values described above, we perform a non linear fit of such data using Eq.(\ref{ede11}) and $\Omega_*$ as the only free parameter. The best fit is found for 
\begin{align}\label{bestfit}
\Omega_*=0.280 \pm 0.117\;,
\end{align}
which implies $w_{de}=-0.867 \pm 0.056$. We also specify that we obtain $j_0=0.875$, thus introducing a discrepancy of only about $12\%$ with respect to a base $\Lambda$CDM picture at very low $z$. For comparison with respect to our model (from now dubbed EDE), we introduce here the power-law profile $\mathcal{H}_0(z)=H_0 (1+z)^{-\alpha}$ with the obtained best fit $\alpha=0.016\pm0.009$ (dubbed as PL model) from \cite{Dainotti2021apj-powerlaw}. We also remark that, in the following, when we refer to $\Lambda$CDM model, we intend a constant effective Hubble parameter $\mathcal{H}_0(z)=H_0=73.5$. In Fig.\ref{fig-H0}, we plot the profiles of $\mathcal{H}_0(z)$ for the three distinct models together with the considered data points. In Fig.\ref{fig-Omega-de}, we instead present the profile of the obtained $\Omega_{de}(z)$ together with the evolutive dark energy (expressed in the standard $w_0w_a$ parametrization) constrained by DESI \cite{DESI:2024mwx} (i.e. with $w_0=-0.82\pm0.063$ and $w_a=-0.75^{+0.29}_{-0.25}$, for the plot we take the central values).
In order to better compare our model to the DESI $w_0w_a$CDM model, we introduce an effective equation of state parameter $w_{e\!f\!f}$, that, from Eq.(\ref{ede4}) and the definitions above, reads as follows: 
\begin{equation}
w_{e\!f\!f}(z) \equiv w_{de} - \frac{3H \bar{\xi}}{\rho_{de}} = 
w_{de} - \frac{\Omega_*}{3\Omega_{de}} \sqrt{\Omega_{m0}(1+z)^3 + \Omega_{de}} 
\, , 
\label{weff}
\end{equation}
The plot of this $w_{e\!f\!f}$ as a function of the redshift is in Fig.\ref{fig-wz}, where we also overplot the profile of the DESI counterpart, i.e. $w(z)=w_0+w_a z/(1+z)$. However, because of the condition on $q_0$ to take the $\Lambda$CDM value, our $w_{de}$ value differs from $w_{e\!f\!f}(z=0)=-1$. It is remarkable that, while $w_{de}>-1$, the bulk viscosity effect provides $w_{e\!f\!f}\leq-1$.

Concerning the statistical relevance of our model, we obtain $\chi_{\text{red}}^2=2.066 \text{(PL)}, 2.095 \text{(EDE)}, 2.176 (\Lambda\text{CDM})$. While none of the models achieves no too optimized fit, our model exhibits improved overall performance against to the standard $\Lambda$CDM framework and performs on par with the PL model. The dense sampling at low redshift and the substantial degrees of freedom result in (inverse) p-values that are nearly $1$ for all three cases. In fact, the ordering of $\chi_{\text{red}}^2$ values highlights a slightly PL model's superior significance in fitting the binned data. This model was derived from the hypothesis that SNe might be affected by an intrinsic redshift evolution \cite{2021ApJ...914L..40D}. Moreover, a natural explanation for the PL variation of the Hubble constant has been proposed in \cite{schiavone_mnras}, where a specific (viable) metric in $f(R)$ gravity, viewed in the Jordan frame, was conjectured. Moreover, we report the following values for standard AIC and BIC analysis: $AIC=103.12 \text{(PL)}, 103.17 \text{(EDE)}, 104.11 (\Lambda\text{CDM})$ and $BIC=104.8 \text{(PL)}, 104.86 \text{(EDE)}, 104.11 (\Lambda\text{CDM})$. While the AIC method still confirms that our model is statistically preferable with respect to the $\Lambda$CDM, the BIC analysis suggests that the models are not distinguishable statistically.
In fact, considering EDE as our reference model, denoted with ``0'' and the $\Lambda\text{CDM}$ and PL models as the other two ones, denoted with ``1'' and ``2'', respectively, we here use the definition of $2 \log_e B_{1,2;0}= BIC_0-BIC_{1,2}=\Delta_{BIC}$ where $B_{1,2;0}=e^{(BIC_0-BIC_{1,2})/2}$. Thus, we obtain $B_{10}= 1.03$ and $B_{20}= 1.46$. Given the current values, it is hard to distinguish statistically which model is the most favored. We here have referred to the table in \cite{Kass01061995}. However, we pinpoint that the BIC works assuming that the variables are independent. Since SNe Ia are correlated, the BIC test may less trustworthy. Thus, we prefer to rely on the AIC, since it does not include the number of SNe Ia. Indeed, it is also worth noting that some of the SNe Ia are repeated in the sample, since they belong to different survey. We conclude, by specifying that the last point at high $z$ values has a larger y-errorbar compared to the other previous bins. Nevertheless, in our previous analysis \cite{Dainotti2021apj-powerlaw} we have shown that this trend exist in 3, 4, 20 and 40 bins and an incoming analysis using equivolume binning (Dainotti et al. in preparation) confirms the decreasing trend of $H_0$ regardless of the bin numbers and the binning procedure. Moreover, given the paucity of data points and the statistics within each bin, performing MCMC analysis on $\Omega_*$ will not allow to show if the trend exists within 2 sigma. More data are needed to verify such a trend beyond 2 $\sigma$.

Summarizing, we underline how the current evolutionary dark energy model offers a more compelling fit to the data compared to the base $\Lambda$CDM model. This result not only points to the existence of a running Hubble constant with redshift across the Pantheon sample, but also suggests that the observed deviation from the expected behavior of the Hubble parameter can be well explained by an evolutionary dark energy term. This statistical evidence strongly supports the idea that the vacuum energy density must vary with increasing redshift, in agreement with the conclusions of the DESI Collaboration. Our analysis also highlights the importance of extending this work to the Pantheon+ sample \cite{Brout:2022vxf}. The enriched sample of SN events in Pantheon+ could offer improved statistical accuracy for binned analyses. Achieving this will require a substantial data analysis effort, marking a key direction for future studies. In fact, we are working on a different sampling strategy \cite{dainotti_in_prep}, where we remove duplicates according to different criteria and combine also different samples. The purpose of using here the Pantheon sample is to compare with similar analysis that has been done in our past work.
A future work will also investigate the Pantheon+ sample and different statistical assumptions as done in \cite{Dainotti:2023ebr}.  We can conclude that the proposed non-equilibrium evolution of dark energy components presents a viable scenario for the late-time dynamics of the Universe, and it is preferable over the Standard Cosmological Model as well as other possible astrophysical or modified gravity frameworks that have been previously explored.


\end{document}